\documentclass{article}

\usepackage[preprint]{neurips_2026}
\usepackage{etoolbox}
\usepackage{makecell}
\usepackage{url}
\expandafter\def\expandafter\url\UrlBreaks{\UrlOrds\do\/\do\-\do\.\do\?\do\=\do\&\do\_}

\makeatletter
\def\@noticestring{}  % Removes the "Preprint" text
\patchcmd{\@notice}{\enlargethispage{2\baselineskip}}{}{}{} % Removes extra page stretching
\makeatother

\usepackage[utf8]{inputenc}
\usepackage[T1]{fontenc}
\usepackage{hyperref}
\usepackage{booktabs}
\usepackage{amsfonts}
\usepackage{nicefrac}
\usepackage{microtype}
\usepackage{xcolor}

\title{The Biomimetic Architecture of Software 4.0}

\author{
  Philip Sheldrake \\
  Unnamed Labs \\
  Amsterdam \\
  \texttt{me@philipsheldrake.com}
  \And
  Dirk Scheffler \\
  Unnamed Labs \\
  Karlsruhe \\
  }

\begin{document}

\maketitle

\begin{abstract}

Dominant programming paradigms inherit an execution model optimised for a bygone era of a single human mind instructing a local machine, leaving contemporary systems burdened with path dependencies. When forced to host multi-dimensional, connectionist intelligence, this brittle compositional model breaks down: the probabilistic and symbolic worlds do not interface cleanly. While contemporary Software 3.x frameworks attempt to patch the mismatch by encasing large language models (LLMs) in increasingly complicated harnesses, this spiralling architectural complexity only compounds the carrying cost of static code assembly. To address the cause rather than the effects, this paper introduces Software 4.0 --- an autopoietic heterarchy of human intelligence, neural AI, and natively reflective symbolic substrate. At its core is a simple premise: intelligence survives its ignorance by giving the unknown a form it can keep and act upon without understanding. Under this paradigm, software is transformed from an inert corpus to be parsed into a self-regulating metabolic network that natively verifies, modifies, and evolves its own structural integrity. We present Varela, the programming language that materialises this architecture. By offloading the burden of structural verification to the deterministic substrate, it unlocks a superior inference-time scaling regime --- one where connectionist compute translates entirely into deep semantic exploration and hypothesis traversal rather than the steep computational and financial cost of simulating structural constraints probabilistically. Moving beyond the `Software Factory' mindset, we outline the foundations required to ground connectionist intent in deterministic structure.

This is a foundational vision paper; formal specification and empirical evaluation of Varela are ongoing.

\end{abstract} 
\section{Introduction: Exo-intelligence}
Most dominant programming paradigms in widespread use today were conceived and substantially developed before the advent of the World Wide Web, let alone large language models (LLMs). The paradigms --- ranging from early assembly to contemporary memory-safe languages --- share a siloed execution model designed for a single human mind to instruct a local machine.

While non-human co-authorship is currently practised within these paradigms (as seen in the current Software 3.x eras, see Table \ref{taxonomy}), the interaction is fundamentally brittle.

For example, models show severe performance degradation under minor prompt perturbations \cite{Rabbi2025Robustness}, autonomous agent trajectories suffer cascading failures  \cite{Jiang_2026}, and surface-level text fluency decouples from functional correctness \cite{He2026Bridging}.

These legacy paradigms lack what we call endo- and exo-homoiconicity. Homoiconicity here deepens the classical homoiconic properties found in languages such as Lisp. Rather than merely equating source text with data structures, it establishes a strict structural isomorphism across the full representational substrate --- unifying types, invariants, and execution paths into a canonical form through which data, configurations, function calls, errors, and behavioural contracts are natively composed, evaluated, and resolved. Everything is a model, and models modify models.

Exo-homoiconicity unifies communication and reflection. It is an operational state where types and properties natively declare their own capabilities and structural affordances, allowing external intelligences, human or machine, to read the system's structure directly without a translation layer. Interoperability ceases to be an exercise in schema mapping and becomes an act of native structural intelligibility. It frames reflection as the continuity of meaning across incomplete understanding: an agent that does not fully grasp a structure can still act on it correctly, because the structure declares its own affordances and enforces its own boundaries.

Exo-homoiconicity is the essential means to a critical end: exo-intelligence, which we define as the capacity to externalise and internalise knowledge. Whereas endo-intelligence represents the bounded intelligence of internal data and code organisation, exo-intelligence is a substrate-level capacity that enables domain-agnostic intelligibility across heterogeneous intelligence modalities without bespoke integration or prior schema alignment.

Exo-intelligence is distinct from classical semantic interoperability, ontology sharing (e.g., OWL), and linked data paradigms. Where legacy semantic frameworks rely on static, schema-bound taxonomies requiring \textit{a priori} alignment to resolve meaning, an exo-intelligent system derives its intelligibility directly from its endo- and exo-homoiconic architecture.

Two traditions shape our design: mathematical self-reference and biological self-organisation. We introduce supersoftware as their embodiment.

\section{The Evolution of Software Paradigms: The Strange Loop}
The classical paradigm of human-authored, deterministic source code defines Software 1.0. Deep neural networks first prompted the hypothesis that learned weights would replace explicit logic entirely, a paradigm termed Software 2.0 \cite{karpathy2017software}. Rather than total displacement, however, the evolutionary trajectory instead yielded a hybrid configuration: the systematic delegation of deterministic code generation to probabilistic language models, dubbed Software 3.0 \cite{karpathy2025software}. Within the 3.x continuum, we identify a rapid architectural progression through three distinct eras (see Table \ref{taxonomy}).

The baseline Software 3.0 can be understood as episodic, decoupled co-authorship, in which the model functions primarily as an unstructured autocomplete. The developer prompts for discrete procedural blocks and evaluates the resulting outputs largely through local inspection and testing rather than through integrated reasoning or formal guarantees. At its extreme, where developers rely heavily on generated code without fully understanding its implementation, this practice has become known as \textit{vibe coding}. In such workflows, responsibility for establishing correctness remains primarily with the human developer through post hoc verification.

Recognising that unconstrained token generation is inherently fragile, Software 3.1 introduced closed-loop execution feedback. Language models are programmatically coupled with local testing environments. Upon generating a candidate code block, the substrate evaluates the program automatically; runtime failures or compiler exceptions generate explicit error streams that are piped directly back into the model's context window. This automates a reactive, iterative self-correction loop, though the feedback is only as good as the error messages: shallow signals in, shallow corrections out.

The contemporary state of the art, Software 3.2, attempts to scale these dynamics to complex macro-level software architectures through sophisticated meta-frameworks. As extensive codebases exceed the processing limits of singular model context windows, the connectionist core must be encased within an elaborate programmatic architecture. This era relies on two primary mitigations: first, intensive context engineering \cite{mei2025survey}, which systematically primes the model's working memory with curated repository metadata and static schemas prior to generation; and second, the construction of multi-agent pipelines governed by a programmable harness \cite{khattab2023dspy}. Here, heterogeneous models engage in a multi-stage, iterative relay --- alternately generating, reviewing, and isolating code execution paths until the framework forces a compiled output.

The structural fragility of this ungrounded paradigm reduces the engineering discipline to a form of techno-magical thinking \cite{evans1937witchcraft, dingemanse2026}. A stark manifestation of this occurred in early 2026 with the release of Claude Code, a state-of-the-art agentic development tool with runtime execution capabilities \cite{townsend2026}. Despite comprising over 519,000 lines of TypeScript architecture, its operational loops rely heavily on a probabilistic foundation rather than a deterministic one \cite{townsend2026}. Lacking a native symbolic substrate to enforce structural invariants, its creators resorted to embedding natural language prompts, explicitly pleading with the model to ``report outcomes faithfully'', ``never characterize incomplete or broken work as done'', and ``be careful not to introduce security vulnerabilities'' \cite{anthropic2026}.

Software 4.0 emerges as the direct architectural resolution to this impasse, defined here as the autopoietic heterarchy of human intelligence, neural AI, and a reflective, structurally verifiable symbolic substrate.

Autopoiesis describes self-producing and self-maintaining systems \cite{maturana1980} (see Section~\ref{sec:selfmodels}). A heterarchy is an organisational form whose elements hold no fixed rank, where which element governs which depends on context and relation \cite{mcculloch1945}. In Software 4.0, this heterarchy is animated by a \emph{strange loop} --- Douglas Hofstadter's term for a structure in which traversing the levels of an apparent hierarchy reveals that top and bottom are locked in a cycle of mutual creation \cite{hofstadter1979}. Escher's \emph{Drawing Hands} (1948) is the canonical image: each hand draws the other into existence, dissolving the spatial metaphors of ``up'' and ``down''. In the same way, human intelligences, neural AI, and a reflective symbolic substrate do not sit in a fixed chain of command; each continuously shapes, constrains, and actuates the others.

This reflective symbolic substrate makes the strange loop tractable, closing the semantic void (see Section~\ref{sec:semanticvoid}) and anchoring Software 4.0 --- which we embody as \emph{supersoftware}. We use \emph{substrate} in its biological sense as the active medium within which coupled processes enact themselves rather than a passive foundation beneath the loop, while welcoming its familiarity to computer scientists as a structural execution layer. It provides the deterministic structural integrity and verifiable state through which human grounding and neural ``reckoning'' cohere into systemic ``judgement'' \cite{smith2019} (see Section~\ref{sec:judgement}), while sustaining the generative capacity to form and evolve self-reproducing structures.

Throughout Software 3.x, the intelligence in the loop is the LLM's, constrained from outside by a harness that contributes none of its own. In Software 4.0, intelligence is realised by the heterarchy itself: human intent, neural proposal, and symbolic reflection and verification constitute one another within the strange loop. The structural form of intelligence is no longer a controller wrapped around a model.

\begin{table}[t]
  \caption{Comparative taxonomy of software paradigms}
  \label{taxonomy}
  \centering
  \renewcommand{\arraystretch}{1.75}
  \begin{tabular}{llll}
    \toprule
    \textbf{Era} & \textbf{Paradigm} & \textbf{Modality} & \textbf{Structural Form of Intelligence}\\
    \midrule
    1.0   & traditional software     & instruction     & source code \\
    2.0 & neural networks          & parameter       & weights \\
    3.0& programmable LLMs        & prompts         & weights + context window \\
    3.1   & agentic engineering      & agentic loops   & weights + harness \\
    3.2   & \makecell[tl]{compiled agentic\\ engineering} & \makecell[tl]{optimised\\ pipeline} & \makecell[tl]{weights + instrumented context\\ + programmatic harness}\\
    4.0   & supersoftware            & strange loop    & heterarchy (autopoietic closure)\\
    \bottomrule
  \end{tabular}
\end{table}

\section{The Semantic Void: The Limits of Software 3.x}\label{sec:semanticvoid}
Our use of semantic here refers to the operational semantics: the strict, formal invariants, execution pathways, and state dependencies natively enforced by the symbolic substrate itself.

Current agentic workflows operate within a structural blind spot. Even when confronting modern, well-factored codebases designed for automated consumption, an LLM interfaces with a fragmented landscape of static file artefacts, frozen dependencies, and passive documentation. As traditional codebases exist as dead text strings until batch compilation or interpretation, an external model must compensate probabilistically for an execution context it cannot natively see.

This blind spot cannot be bypassed by merely exposing or training models on compiler Intermediate Representations (IR), such as Meta's LLM Compiler \cite{cummins2024meta}. This model still requires a rigid, external software harness to prevent it from hallucinating invalid states, even after feeding a transformer 546~billion tokens of pure compiler IR and assembly. Because an IR remains a static serialisation of code syntax, an external model is still forced to probabilistically simulate an execution trace over inert artefacts. No amount of inference on traditional software code or IR can reconstruct what was never made explicit. LLMs are, in effect, generating text into a void.

Not only does this structural decoupling create engineering debt, it also introduces significant socio-technical friction. As non-human co-authorship scales through external interfaces, the human role shifts from creative design to rapid, reactive verification --- a phenomenon sometimes referred to as \textit{pull request fatigue}. A cognitive misalignment leaves developers trapped in an unsustainable cycle of monitoring automated intent \cite{coderabbit2025state}. Long-horizon agent runs remain unpredictable, failing in cascades that are hard to diagnose or prevent \cite{hagele2026hotmessaidoes}.

Such structures expose a profound probabilistic-symbolic impedance mismatch between traditional software architectures and connectionist intelligences. Where legacy computing environments expect two deterministic, precise systems to interface cleanly, the integration of an LLM introduces fundamental paradigmatic frictions: \textit{determinism} (fuzzy token vectors vs. invariant compiler correctness), \textit{state management} (stateless, sliding context windows vs. rigid runtimes), \textit{type safety} (probabilistic token streams vs. static structural boundaries), \textit{cost and latency} (expensive matrix multiplication vs. $O(1)$ algorithmic checks), and \textit{debugging} (non-deterministic hallucinations vs. deterministic execution traces). 

Where Software 3.x attempts to patch the impedance mismatch by wrapping the system in increasingly complicated harnessing, Software 4.0 collapses the boundary. It embeds a reflective substrate that binds probabilistic intent to deterministic invariants natively at the exact point of generation. To operationalise this native binding, our programming language Varela introduces a qualitative shift toward epistemic late binding. The substrate's grammar defines a fixed envelope of types, constraints, and dependencies, while leaving the concrete execution path open. That path is decided only upon interaction with a reasoning agent, and every such interaction is verified against the envelope before it takes effect.

\section{Foundational Mechanics: Parsimony and Reflection}
We inherit Ada Lovelace’s insight that symbolic enumeration grants a machine domain-agnostic universality \cite{IEEE2022}, a concept then formalised mathematically by Gödelian numbering, which demonstrates that formal syntax, proofs, and computations can be cleanly mapped as invariant structural primitives within a single numeric space \cite{smullyan1992godel}.

Software 4.0 couples such structural parsimony with computational reflection --- the capacity of a system to evaluate and manipulate its own internal state recursively \cite{smith1982reflection}. By executing these reflective, homoiconic dynamics within a reduced type landscape, the platform directly honours the core ethos of the RISC heritage --- stripping away arbitrary syntactic abstraction layers to yield optimal execution performance and systemic resilience \cite{pattersonditzel1980}. This cross-stack mathematical and structural parsimony acts as the primary defence against structural entropy --- the unchecked growth of structural degrees of freedom and resulting loss of coherence.

\section{Foundational Cybernetics: Model Drift and Variety}
Cybernetics bridges engineered, artificial systems, and evolved, natural systems such as organisms and societies \cite{heyligen2003}. It ``approaches systems in terms of purpose --- the purpose attributed 
to the system by observers observing `the system' as well as our own purpose as observers in explaining `the system'.'' \cite{labcybernetics2026}

At the core of this synthesis is endogenous control: the compensation of environmental perturbations must be embedded in the system's own dynamics. In our contexts, while an external harness can execute high-frequency feedback loops to enforce operational boundaries, it remains exogenous and so decoupled from the internal state-space of the system it observes. Because an external controller merely runs a lagged model of the system --- rather than regulation emerging from logic distributed within and co-evolving with the substrate --- its representation inevitably falls out of step with the reality it intends to govern.

This vulnerability is compounded because a system doesn't just fall into drift through exogenous isolation, but also through insufficient regulatory variety. Satisfying Ashby's Law requires that a regulator maintains control only to the degree that its internal variety matches or exceeds the system's variety \cite{ashby1956}. Whereas an external harness has access only to a system's observable outputs and a limited set of interventions, structurally bounding its variety by the narrower bandwidth of the orchestrator-substrate interface, a self-regulating system has direct, unmediated access to its own internal states. Its regulatory variety is thus structurally unbounded by an external boundary layer.

It follows that externalised harness architecture can never achieve the regulatory capacity of a constitutively self-modelling substrate, regardless of sophistication.

When a system entails a direct, endogenous control mechanism, it achieves a reflexivity concerning its own operations --- the capacity to observe and describe itself as part of the system it regulates \cite{vonfoerster2003}. This reflexive closure, central to second-order cybernetics, enables a far more complex capacity for systemic sense-making --- how an autonomous organism appreciates a subset of its environmental couplings as meaningful \cite{varela2017embodied} --- and, by extension, provides the structural foundation for shared, intersubjective meaning-making across the human-machine boundary \cite{deacon2011incomplete}.

Nature, which could never rely on external harnessing, solved this problem long ago.

\section{Foundational Biological Principles: Constitutive Coupling}
Computer science repeatedly rediscovers or mimics mechanisms long realised in biological systems, suggesting not occasional analogy but structural convergence driven by shared constraints on information processing under noise, decentralisation, and limited resources.

This is evident in neural computation models of memory and learning, from Hopfield networks \cite{hopfield1982neural} to backpropagation in distributed connectionist systems \cite{rumelhart1986learning} --- the latter being functionally analogous to biological credit assignment even if mechanistically distinct. It is equally explicit in evolutionary search paradigms such as genetic algorithms \cite{holland1975adaptation}, and in distributed coordination, swarm and collective intelligence methods, ranging from flocking models \cite{reynolds1987flocks} to ant colony optimisation \cite{dorigo2004ant}. Even foundational network protocols such as TCP congestion control \cite{jacobson1988congestion} embody homeostatic feedback principles closely analogous to biological regulation under resource constraints. Taken together, these results reinforce the perspective that modern computation does not merely draw inspiration from biology, but repeatedly converges upon its functional strategies when confronted with equivalent systemic constraints.

Biological systems demonstrate that resilient configurations persist across evolutionary timescales through constitutive coupling: where distinct roles are so operationally interdependent that none can function in isolation \cite{kirschner2005, wagner_robustness_2005, kitano_biological_2004}. This relational architecture reflects Rosen’s formalisation of organisms as being `closed to efficient causation' --- a state where every functional role is produced and maintained by the internalised causal loops of the system itself \cite{rosen1991life}.

Software 3.x couples probabilistic generation to code without a constitutively coupled structural anchor. Varela supplies that anchor: a homoiconic substrate that bounds neural reasoning with verifiable invariants, the role organic code plays in the cell.

We are guided by four generative axes:

\subsection{Axis 1: Encoding and Expression}
The distinction between information storage and functional execution, exemplified by DNA and protein systems, is associated with biological processes that maintain stability while remaining highly responsive to context. The pattern recurs across lineages: coupling that is permissive in context but constrained in structure \cite{maturana1980, alberts2015, ptashne2002genes}.

Varela takes the encoding role, replacing opaque encapsulation with transparent reflection and epistemic late binding. Supersoftware is the expression, with model definitions interpreted in the moment, in context.

\subsection{Axis 2: Novel Exploration and Low-Variance Execution}
The distinction between flexible, generative computation and low-variance predictive execution, exemplified by the cortical and cerebellar systems, enables neural processes to explore novel strategies while maintaining reliable performance through collaborative and corrective coupling. The cerebellum’s forward models bias cortical proposals \cite{wolpert1998}.

In the contexts here, the LLM generates fluid stochastic creativity and probabilistic inference, expressively bounded by supersoftware's structural demands. Supersoftware guarantees repeatable, structurally anchored execution, constraining the LLM’s generative volatility. The coupling is collaborative and corrective: the model operates within an uncompromising reality where its proposals are immediately forced into structural reconciliation.

\subsection{Axis 3: Intent and Self-Regulation}
The distinction between intentional, goal-directed behaviour and involuntary, anticipatory self-regulation, exemplified by somatic and autonomic nervous systems, enables organisms to pursue adaptive goals while maintaining internal stability through allostatic, purposive coupling (where the autonomic system predicts and pre-adjusts set points) \cite{sterling2012allostasis}.

In Varela, human intelligences provide the authoritative intent, directing without writing lines of code. Contextual interfaces assist the systemic understanding needed to establish governing structural invariants (moldable, live visual environments tailored to domain state, e.g., \cite{gtoolkit2026}), and LLMs interpret human direction to generate fluid, operational sub-intents that remain structurally bounded by the substrate.

LLMs also participate in the self-regulation loop. They make proposals, receive structural feedback, and iterate, collaborating with supersoftware toward coherence without needing human intervention at every step. The supersoftware embodies structural self-regulation, self-verification, self-correcting execution. It grounds and verifies LLM proposals, coupled to human intent through purpose rather than instruction.

\subsection{Axis 4: Diverse Generation and Selective Retention}
The distinction between combinatorially diverse generation and selective, affinity-based retention, exemplified by the immune repertoire’s diversity and selection for stronger antibody binding, enables the system to respond to novel threats and not attack itself \cite{burnet1957modification}. In other words, retention emerges only from successful discrimination, and discrimination is sharpened by immunological memory.

In our contexts, human and LLM intelligences generate diverse proposals (code, structures, natural language translations), and supersoftware provides the structurally verified identity (a sense of `self'). It discriminates valid from invalid, and retains successful patterns as `verified memory' instantiated as a distributed registry of structural invariants and successful homeostatic adaptations accumulated across network deployments.

\subsection{Self Models}\label{sec:selfmodels}
Such constitutively coupled functional differentiations encompass the system’s ability to maintain a model of itself without a central architect --- to encode, inspect, and act on its own structure \cite{rosen1985anticipatory}. For our four axes, the underlying dynamics are modelled in terms of autopoiesis, encephalic active inference, allostasis, and immunological memory. While these four distinct regimes operate heterarchically, autopoiesis acts as our primary ontological guide across this thesis and taxonomy. Autopoietic closure is the foundational precondition for the remaining three axes, providing the bounded, self-producing structural identity --- the system's operational perimeter --- without which it cannot execute allostatic regulation, engage in predictive active inference, or leverage immunological memory.

The axes signpost software engineering's movement from mechanistic assembly toward an ecology of self-referential differentiation. The next section synthesises this movement.

\section{Synthesis: The Architecture of Software 4.0}

\subsection{The Self-Hosting Substrate}
For software to operate as an autopoietic medium rather than a static artefact assembled from outside, its language cannot sit apart from the structures it governs. Any Software 4.0 language must therefore be self-describing and normalised at the language level: its definitions, types, constraints, and operational state must inhabit the same reflective substrate as the domain systems expressed in it. If a language relies on an external host or detached meta-layer, structural verification and endogenous adaptation collapse.

Varela satisfies this architectural requirement directly. Varela is itself expressed in Varela: the language is a supersoftware system, self-describing and normalised at the language level. A domain system expressed in Varela extends the substrate rather than stacking a second layer on top.

\subsection{Coupling LLMs and Supersoftware}\label{sec:coupling}
Recent work in code language modelling suggests that programming languages with highly explicit and regular structural representations should, in principle, align well with the inductive biases of transformer architectures. Studies on AST/CST-aware learning have repeatedly shown that models benefit when syntactic structure is made explicit rather than inferred indirectly from surface token streams \cite{agarwal2024}. Likewise, syntax-aware training strategies show that incorporating structural information materially improves code understanding and generation \cite{gong2025}. Homoiconic languages are notable in this context because their surface syntax already approximates a serialised abstract syntax tree, meaning the distance between program text and program structure is uniquely compressed.

This optimisation is strongly validated by connectionist research into meta-learning. Lampinen and McClelland (2020) demonstrated that when a neural architecture exploits a deep functional analogy between basic tasks and higher-order task transformations --- mapping both into a single, shared representational space via a Homoiconic Meta-Mapping (HoMM) framework --- it achieves remarkable cognitive flexibility \cite{lampinen2020transforming}. By unifying mapping within this homoiconic space, the network frequently reaches 80\%~to~90\% performance on novel, zero-shot task adaptations, significantly outperforming alternative connectionist baselines where task context is merely appended as an external input to a standard feed-forward network. Software 4.0 satisfies this structural constraint natively. By replacing text-mediated interfaces with a unified, reflective topology where code, state, and domain context share an identical structural presentation, the model develops a deep structural grasp of underlying logic, radically improving out-of-distribution generalisation and correctness.

Contemporary connectionist systems demonstrate architectural fragility in practice. Structural perturbation studies confirm that transformer models are disproportionately sensitive to syntactic and compositional irregularities; disrupting structural coherence or introducing minor surface-level prompt perturbations impairs model reasoning far more severely than semantic alterations \cite{waheed2025, Rabbi2025Robustness}. This suggests the models lean heavily on token-level patterns to track structure.

To mitigate this dependency, structural parsimony and homoiconic representations become an architectural necessity rather than a stylistic preference. By minimising parser ambiguity, eliminating heterogeneous syntactic forms, and avoiding operator precedence irregularities, such representations drastically reduce syntactic entropy. In combination with an unambiguous grammar design, this structural regularity minimises the cognitive overhead of the text layer, offering an optimal target for both pattern learnability and downstream synthesis fidelity.

Empirical evidence confirms that code-centric reasoning trajectories yield significantly higher performance than natural language alone, validating the thesis that connectionist systems optimise when navigating strict structural environments rather than unstructured text \cite{hwang2026}. Software 4.0 takes this insight to its logical conclusion. Rather than just using code-centric representations as an optimised bridge, the neural-symbolic gap is collapsed into a natively self-modelling substrate. By codifying the world model into the substrate itself, the architecture no longer estimates structure; structure is a given.

Admittedly, current baseline datasets do not yet provide direct empirical isolation of these homoiconic advantages over conventional languages such as Python or Java. Because existing coding benchmarks overwhelmingly under-represent homoiconic ecosystems, controlled comparisons remain difficult, and corpus-scale effects presently dominate observed language-specific performance differences. However, this empirical deficit is mitigated if one adopts target-driven training strategies designed to mitigate dependence on massive naturally occurring corpora --- for example, through structure-aware pre-training, synthetic curriculum generation, or explicit AST-level supervision. Critically, because a homoiconic substrate minimises syntactic entropy by design, these structure-centric methods can be executed at a materially lower computational and financial cost than for conventional, high-entropy languages.

This neurosymbolic coupling redefines the scaling frontier that ungrounded neural progression cannot cross. As connectionist models lean heavily into inference-time scaling regimes, attempting to maintain structural coherence purely within probabilistic search trajectories imposes an unsustainable computational and financial penalty. In ungrounded architectures, the model must expend a significant portion of its search budget and autoregressive token overhead simply to simulate structural constraints, track state boundaries, and enforce type safety rules. By contrast, shifting these invariants to a reflective, symbolic substrate offloads the entire burden of structural verification to a deterministic engine. Consequently, the Software 4.0 paradigm unlocks a qualitatively superior inference-time scaling regime --- one where compute translates entirely into semantic exploration and deep hypothesis traversal rather than superficial structural simulation.

To guarantee this structural transparency, Varela operates as a constitutive knowledge representation and reasoning (KR\&R) system \cite{brachman2004knowledge} paired with logic-programming-style structural pattern matching and invariant enforcement \cite{kowalski1979algorithm}. While it inherits the KR\&R tradition's commitment to explicit structured representation and the logic programming tradition's commitment to declarative structural invariants, it departs from both by operating in a reactive rather than batch mode, embedded directly as the execution substrate rather than applied as an external abstraction layer. Traditional KR\&R models are descriptive --- they build passive, abstract models of things external to the software. In Software 4.0, the world model is constitutive; it is the software itself.

This architecture functions essentially as a computational ribosome \cite{barbieri2015code} --- an organic code layer that projects the live, endo- and exo-homoiconic state (comprising active types, properties, constraints, and dependency graphs) directly into the model's context. Rather than parsing dead text strings to infer system state, the LLM enacts a live, self-describing model of the system, with its invariants checked natively in real time.

While humans may no longer author this substrate directly --- engaging instead through natural language and contextual interfaces --- the supersoftware remains conceptually legible. It provides the ground truth that allows the neurosymbolic AI to explain itself, humans to understand it, and the system to maintain its own structural integrity and identity. In other words, legibility is not only about verification; it is what makes the system's ongoing reformation accountable (see Section~\ref{sec:judgement}).

\subsection{The `Narrow Waist' of Architectural Intent}
In systems architecture, a narrow waist refers to a minimal, universal interface that allows diverse higher-level applications to run on diverse lower-level infrastructures. In traditional compiler design (e.g. LLVM), an intermediate representation (IR) acts as the narrow waist between many front-end languages (e.g. C++, Rust, Swift) and various back-end architectures (e.g. x86, ARM, RISC-V), preserving operational semantics while abstracting implementation details.

The LLM industry's current proxy for a narrow waist --- Model Context Protocol (MCP) --- standardises contextual transport and API schemas \cite{mcp2024}. It addresses interoperability in Software 3.x by defining a uniform protocol for model-tool communication. It standardises how models request and receive context or tool execution results via structured message exchanges. However, it does not define or constrain the internal execution semantics of tools, which remain opaque to the protocol. MCP governs interaction at the interface boundary, not runtime behaviour within tool implementations.

Varela moves the waist higher by one full level of abstraction from the IR, establishing it directly at the active representational layer to preserve architectural logic, not just execution logic. This isn't just operational semantics, but \textit{meaning} in its formal sense: the purposes and constraints that define the system. Meaning is lost twice when an LLM writes Java and it is compiled to an intermediate representation: first when intent is flattened into static source text, and again when that text is lowered into low-level IR. When an LLM co-authors Varela, the meaning is codified natively. Where MCP uses schemas to standardise how models invoke tools and receive structured results, Varela embeds the living operational invariants, types, and logic dependencies directly into the computational environment, ensuring that the LLM navigates an uncompromising symbolic reality rather than an ungrounded API interface.

Akin to the ribosome, the substrate's homoiconicity affords a far greater application of domain-agnostic algorithms. Guided purely by declared schema without semantic awareness, they can traverse, validate, transform, merge, patch, serialise, replay, or persist any data form, analyse relationships, track causality, and preserve referential integrity.

Within the Varela architecture, this capability is realised by Varela Exchange — the foundational exchange protocol for structural serialisation and representation, encoding the system's structural affordances directly into the communication layer. Designed to store and transport type-safe context referentially, comparably, and polymorphically, Varela Exchange enables deeply nested and highly streamable data structures to cross system boundaries without structural degradation. Its scope extends beyond wire transmission: it unifies functional logic, persistence, reflection, schema metadata, presentation layers, configuration states, and runtime error handling into a single cohesive medium. Consequently, it acts as the common representational plane through which connectionist systems can safely read, mutate, and reason over complex symbolic architectures without intermediate compilation.

\section{On Judgement}\label{sec:judgement}

To satisfy Brian Cantwell Smith's formal distinction between reckoning and judgement \cite{smith2019}, and in contrast to Software 3.x, we ground systemic judgement in architectural accountability to the system's own structural reality.

Software 3.x architectures are confined to reckoning because their symbolic manipulations are decoupled from existential consequence: the system has no mechanism by which it is genuinely beholden to its representations, and the stakes that make failure matter sit entirely outside the architecture.

Software 4.0 operationalises the world model through the autopoietic strange loop. Here, the world model is not a passive data file parsed from the outside, but an active metabolic network of homeostatic boundaries, environmental dependencies, and operational constraints. Human intelligences establish the governing invariants through contextual interfaces, and the substrate's homoiconicity keeps the system's structural potentiality and non-negotiable boundaries directly legible to the neural AI. When a neural agent proposes an action, its probabilistic reckonings are forced into structural reconciliation with the system.

Systemic judgement emerges from the strange loop coupling of authoritative human intent, neural reckoning, and symbolic reflection and structural verification as mutually constitutive differentiations. Confronting every proposed action with this constitutional integrity lifts Software 4.0 from rule-bound calculation to responsible engagement with its world. The integrity comes from the settled form of commitments made by the system's human members, for whom structural failure is felt consequence rather than mere technical error. Judgement is therefore systemic: it arises when human-borne stakes are structurally bound to neural reckoning through a self-verifying substrate.

\section{Ecosystem Adoption}
By design, connectionist-symbolic coupling is native to Varela's architecture from day one. Every interaction injects the live world model (the system's current state, its declared constraints, and its dependency graph) directly into the LLM's context window. Because the substrate self-declares its own structural rules, a highly capable model can reason over and mutate Varela code zero-shot: the system itself provides the context that makes valid generation possible. Generated code is evaluated deterministically against the integral world model prior to execution, so incorrect generations are intercepted algorithmically. The runtime
thereby becomes a continuous, real-time alignment loop in which every structural rejection generates a high-fidelity training signal.

This structured feedback operationalises the alternative training strategies anticipated in Section~\ref{sec:coupling}. Traditional programming languages require LLMs to expend vast parameter capacity simply learning to approximate irregular surface syntax; Varela bypasses this tax. We conjecture that endo- and exo-homoiconicity enable the model to focus cleanly on programmatic intent rather than superficial syntactic correctness \cite{de2025tool}, and offer a more expansive, self-referential topology: a denser, more interconnected web of structural patterns within the model's latent vector space. Taken together, Varela's grammar specification, type system, and core codebase constitute a high-fidelity structural seed, one that is vastly richer in semantic signal than an equivalent corpus of unstructured legacy code.

Importantly, the adoption of Software 4.0 does not demand a monolithic re-architecture of legacy estates. Because the substrate natively exposes its own structural affordances, a single Varela system can dynamically project standard, legacy-compliant interfaces --- including REST APIs, gRPC stubs, message queue schemas, Model Context Protocol (MCP) definitions, and POSIX command-line interfaces. This guarantees drop-in interoperability, allowing organisations to deploy localised autopoietic loops that interface seamlessly with conventional Software 3.x infrastructure.

We conjecture that such a new programming language will not be subject to the barriers that have frustrated the adoption of legacy programming languages. Rather than requiring years of developer habituation and community formation, it becomes an empirical optimisation problem for connectionist systems --- effectively making it the native choice of the models themselves. Once integrated into active operation, the capacity to generate structurally coherent code compounds automatically across every model update, benchmark iteration, and live deployment.

\section{Related Work}

\subsection{Distinction from Classical Reflection and Mirror Models}
Exo-homoiconicity is a structural property where types and properties declare their own operational affordances. It turns reflection from a passive mirror into an active instrument of computation, cognition, and intelligence. Through this architecture, automated systems can perceive and manipulate system state contextually and meaningfully --- establishing the structural prerequisite for exo-intelligence. Serialisation becomes intrinsic, transmission decouples entirely from implementation boundaries, and versioning, patching, and replaying become purely structural operations.

Exo-homoiconicity represents a topological inversion of the classical reflective architectures proposed by Smith \cite{smith1982reflection} and Maes \cite{maes1987concepts}. While the `reflective tower' of 3-Lisp was designed to enable internal metacircularity --- where a system’s internal state is made explicit to its own interpreter for autonomous self-modification --- exo-homoiconicity externalises this reflective interface. It ensures that the system’s ability to `see itself' and adjust its execution trace is simultaneously projected outward, self-declaring its potentiality to external intelligences. Traditional structural reflection often leaves a semantic void where the purpose of a code block remains implicit and context-dependent --- an exo-homoiconic substrate, by contrast, guarantees that types and properties declare their own functional capabilities. This allows external intelligences to reason about architectural intent and form without requiring exhaustive pre-trained domain knowledge or brittle procedural parsing.

This approach further departs from the stratified `mirror models' described by Bracha and Ungar \cite{bracha2004mirrors}. While mirrors decouple metadata to preserve system cleanliness for the human developer, they treat reflection as an optional tool for offline inspection. In Software 4.0, exo-homoiconicity is a constitutive requirement of the substrate rather than a stratified window into a static corpus. It changes the structural form of intelligence from a private, internalised execution loop to a shared, canonical environment, effectively transforming the software from a corpus to be parsed into a coupled substrate to be continuously enacted. The puppet-like manipulation of traditional code is replaced by the autopoietic regulation of a coherent computational organism.

\subsection{Neurosymbolic AI}
Our approach to the structural coupling of intelligences builds upon the foundational work in neural-symbolic integration established by Garcez, Lamb, and Gabbay \cite{d2009neural}. Their systems integrate connectionist learning and symbolic reasoning by translating between two representations: a logic program is compiled into a neural network, trained on examples, and revised knowledge is then extracted back out. This yields an influential training methodology, but one in which the two modalities remain distinct representations coordinated by translation. Software 4.0 collapses this boundary: the neural and symbolic components operate as co-constitutive functional differentiations sharing one representation within a single, closed-loop heterarchy, a coupling that is continuous rather than confined to a training pipeline. The symbolic substrate provides the self-declaring potentiality that neural intelligence actuates, and it is this continuity that moves the architecture beyond iterative communication toward systemic judgement.

Similarly, the neurosymbolic visual question answering (NS-VQA) framework proposed by Mao et al. \cite{mao2019neuro} demonstrates the power of combining neural perception with symbolic logic. However, such systems typically operate as a modular pipeline where neural models serve as a front-end for symbolic execution. Software 4.0 departs from this sequential arrangement; rather than treating the symbolic layer as a downstream processor of neural outputs, connectionist intelligence and the substrate are constitutively coupled as a single, self-referential architecture.

\subsection{Program Synthesis}
Finally, we recognise the program synthesis tradition, notably exemplified by Gulwani \cite{gulwani2010dimensions}, which automates the generation of programs from high-level specifications or examples. While traditional synthesis is typically framed as an episodic search problem --- generating discrete procedural snippets to satisfy a local constraint --- Software 4.0 shifts the focus from the production of dead code to the autopoietic evolution of a living substrate. We move from the episodic search for a functional program toward the continuous self-regulation and adaptation of a reflective, symbolic substrate.

\section{Conclusion}
Nature began with analogue chemistry and evolved digital code in the form of DNA to preserve structure \cite{hoffmeyer1997signs}. Neural AI has followed the inverse trajectory: emerging from digital silicon only to collapse into the statistical, ungrounded representations of connectionist weights. While contemporary architectures attempt to arrest this entropy through inference-time reasoning traces and scaffolding, these connectionist loops remain ungrounded. Lacking a native symbolic substrate, they must expend substantial computational overhead simply to simulate structure probabilistically, leaving the generation trajectory to suffer a fatal drift.

Where conventional software paradigms offer no evolutionary remedy, Software 4.0 resolves this inversion by establishing a framework of computational organic code. Within this reflective substrate, neural execution and symbolic structure are constitutively coupled --- permanently collapsing the divide between fluid probabilistic intent and deterministic structural invariants.

Historically, the rigid design patterns permeating enterprise architectures served primarily to compensate for the expressive limitations of static paradigms. Peter Norvig famously demonstrated that dynamic reflection and homoiconicity render these classical structural workarounds redundant \cite{norvig1996}, yet human practice remained stagnant because the cognitive switching costs outweighed the minor administrative tax of boilerplate. In the age of connectionist synthesis, however, this economic calculus is inverted: while the switching cost trends to zero, the spiralling architectural complexity compounds the carrying cost of legacy paradigms, choking the immense potential of automated generation. Software 4.0 operationalises Norvig's insight and exposes static code assembly as an evolutionary dead-end.

Notably, while the authors of the foundational \textit{Design Patterns} \cite{gamma1995design} text drew early inspiration from Christopher Alexander’s initial work on the synthesis of form \cite{alexander1964notes}, Alexander's own mature philosophy turned decisively away from the mechanical assembly of discrete parts toward the dynamic differentiation of living wholes \cite{alexander2002nature}. Software 4.0 codifies this transition, moving software engineering away from the industrial `Software Factory' mindset characterising Software 3.x and into an autopoietic ecology.

As a foundational vision paper, this work establishes the theoretical architecture for Software 4.0. While the Varela language is in development, its core coordination mechanics have been partially validated within legacy environments. Non-native, locally bounded implementations of these structural constraints have been deployed programmatically within production Java and TypeScript environments, proving the viability of the approach under the rigours of Fortune Global 500 enterprise infrastructure.

This paper establishes the structural prerequisites for individual Software 4.0 systems. The future of the paradigm unfolds along a dual evolutionary horizon. Outwardly, the continuous, decentralised exchange of systemic evolutionary memory points to the emergence of a global computational hyperstructure. Inwardly, cascading these reflective principles through the mechanics of partial evaluation down to the software execution layer allows the system to natively compile its own intent \cite{futamura1971partial}. Both increase the evolutionary velocity of this living software towards multi-scale, self-organising ecologies --- systems structured to cultivate the emergence of collective intelligences and systemic wisdom \cite{johnson2025}.

\bibliographystyle{plain}   % or 'unsrt', 'alpha', 'abbrv'
\bibliography{references}    % matches the .bib filename (no extension)
\end{document}